\documentclass[sigconf,screen]{acmart}

\makeatletter
\def\@ACM@checkaffil{
    \if@ACM@instpresent\else
    \ClassWarningNoLine{\@classname}{No institution present for an affiliation}%
    \fi
    \if@ACM@citypresent\else
    \ClassWarningNoLine{\@classname}{No city present for an affiliation}%
    \fi
    \if@ACM@countrypresent\else
        \ClassWarningNoLine{\@classname}{No country present for an affiliation}%
    \fi
}
\makeatother

\usepackage{amssymb,amsfonts}
\usepackage{ragged2e}
\usepackage{graphicx}
\usepackage{textcomp}
\usepackage{enumitem}
\usepackage[many]{tcolorbox}
\usepackage{soul}
\usepackage{booktabs}
\usepackage{caption}
\usepackage{hyperref}
\usepackage{rotating}
\usepackage{tabularray}
\usepackage{svg}
\usepackage{lipsum}
\usepackage[english]{babel}
\usepackage[autostyle]{csquotes}
\usepackage{pdfpages}
\usepackage{algorithm}
\usepackage{algpseudocode}
\usepackage{longtable}
\usepackage{tabularx}
\usepackage{multirow}
\usepackage{setspace}
\usepackage{makecell}
\usepackage{framed}
\usepackage{booktabs}
\usepackage{balance}

\definecolor{lgreen}{RGB}{50,205,50}

\newcommand\blfootnote[1]{%
  \begingroup
  \renewcommand\thefootnote{}\footnote{#1}%
  \addtocounter{footnote}{-1}%
  \endgroup
}

\AtBeginDocument{%
  \providecommand\BibTeX{{%
    \normalfont B\kern-0.5em{\scshape i\kern-0.25em b}\kern-0.8em\TeX}}}

\title[Assessing AI Detectors in Identifying AI-Generated Code: Implications for Education]{Assessing AI Detectors in Identifying AI-Generated Code: Implications for Education}

\author{Wei Hung Pan}
\email{wpan0017@student.monash.edu }
\affiliation{%
\institution{School of Information Technology, Monash University Malaysia}
\city{Subang Jaya}
\country{Malaysia}}

\author{Ming Jie Chok}
\email{mcho0068@student.monash.edu}
\affiliation{%
\institution{School of Information Technology, Monash University Malaysia}
\city{Subang Jaya}
\country{Malaysia}}

\author{Jonathan Leong Shan Wong}
\email{jwon0100@student.monash.edu}
\affiliation{%
\institution{School of Information Technology, Monash University Malaysia}
\city{Subang Jaya}
\country{Malaysia}}

\author{Yung Xin Shin}
\email{yshi0072@student.monash.edu}
\affiliation{%
\institution{School of Information Technology, Monash University Malaysia}
\city{Subang Jaya}
\country{Malaysia}}

\author{Yeong Shian Poon}
\email{yeon0001@student.monash.edu}
\affiliation{%
\institution{School of Information Technology, Monash University Malaysia}
\city{Subang Jaya}
\country{Malaysia}}

\author{Zhou Yang}
\email{zyang@smu.edu.sg}
\affiliation{%
\institution{School of Computing and Information Systems, Singapore Management University}
\city{Singapore}
\country{Singapore}}

\author{Chun Yong Chong}
\email{chong.chunyong@monash.edu}
\affiliation{%
\institution{School of Information Technology, Monash University Malaysia}
\city{Subang Jaya}
\country{Malaysia}}

\author{David Lo}
\email{davidlo@smu.edu.sg}
\affiliation{%
\institution{School of Computing and Information Systems, Singapore Management University}
\city{Singapore}
\country{Singapore}}

\author{Mei Kuan Lim}
\email{lim.meikuan@monash.edu}
\affiliation{%
\institution{School of Information Technology, Monash University Malaysia}
\city{Subang Jaya}
\country{Malaysia}}

\renewcommand{\shortauthors}{Pan et al.}

\ccsdesc[500]{Social and professional topics~Software engineering education}

\begin{document}

\begin{abstract}
\blfootnote{\textsuperscript{$\ddag$}Zhou Yang is the corresponding author.}Educators are increasingly concerned about the usage of Large Language Models (LLMs) such as ChatGPT in programming education, particularly regarding the potential exploitation of imperfections in Artificial Intelligence Generated Content (AIGC) Detectors for academic misconduct.

In this paper, we present an empirical study where the LLM is examined for its attempts to bypass detection by AIGC Detectors. This is achieved by generating code in response to a given question using different variants. We collected a dataset comprising 5,069 samples, with each sample consisting of a textual description of a coding problem and its corresponding human-written Python solution codes. These samples were obtained from various sources, including 80 from Quescol, 3,264 from Kaggle, and 1,725 from LeetCode. From the dataset, we created 13 sets of code problem variant prompts, which were used to instruct ChatGPT to generate the outputs. Subsequently, we assessed the performance of five AIGC detectors. Our results demonstrate that existing AIGC Detectors perform poorly in distinguishing between human-written code and AI-generated code. 

\end{abstract}

\keywords{Software Engineering Education, AI-Generated Code, AI-Generated Code Detection}
\maketitle

\section{Introduction}
In recent years, an increase in capability has been observed in the development of LLMs, enabling them to generate high-quality, coherent paragraphs, answer questions, and even produce human-like code \cite{naveed2023comprehensive, hou2023large}. LLM are designed to understand and generate human text, and they are trained using vast amounts of data scraped from the internet. Most of these LLMs, developed by major corporations, are generally made accessible to the public, allowing anyone with Internet access to utilize them \cite{chen2023chatgpt}. This accessibility enables individuals to swiftly obtain valuable and reference-worthy answers.

At the educational level, this accessibility can enhance learning and research efficiency while potentially influencing educational assessment and evaluation \cite{susnjak2022chatgpt}. The use of AI-based tools and resources has gradually led to a situation where students increasingly rely on these tools to quickly access answers and information. This growing dependence on AI-driven solutions has a noticeable impact on academic dishonesty \cite{eke2023chatgpt}.

As a consequence, educators find themselves compelled to utilize AIGC Detectors to ascertain whether students are involved in academic dishonesty \cite{uzun2023chatgpt}. While existing AIGC Detectors have proven their proficiency in identifying AI-generated text, their effectiveness in recognizing AI-generated code remains uncertain due to the intricate nature of programming code \cite{wang2023evaluating}. This discrepancy can lead to disparities in the evaluation of students' academic submissions, potentially resulting in unfair grading.

The paper aims to conduct an empirical study to evaluate the performance of different AIGC Detectors in detecting AIGC across diverse contextual and syntactical variations.

Overall, the main contributions of our paper
are summarized as follows: 
\begin{itemize}[leftmargin=*]
    \item We conducted a comprehensive empirical study to assess the performance of five AIGC detectors using 13 variant prompts. To the best of our knowledge, this is the first study specifically evaluating the performance of different AIGC detectors on AIGC generated with various question prompts.
    \item We constructed 13 large datasets, each containing 5,069 samples representing specific code problem prompt variants. Each dataset comprises code problems, human-generated code for that problem, and AI-generated code for the same problem.
\end{itemize}

\begin{figure*}[t]
  \centering
  \tiny
  \includegraphics[scale=0.5]{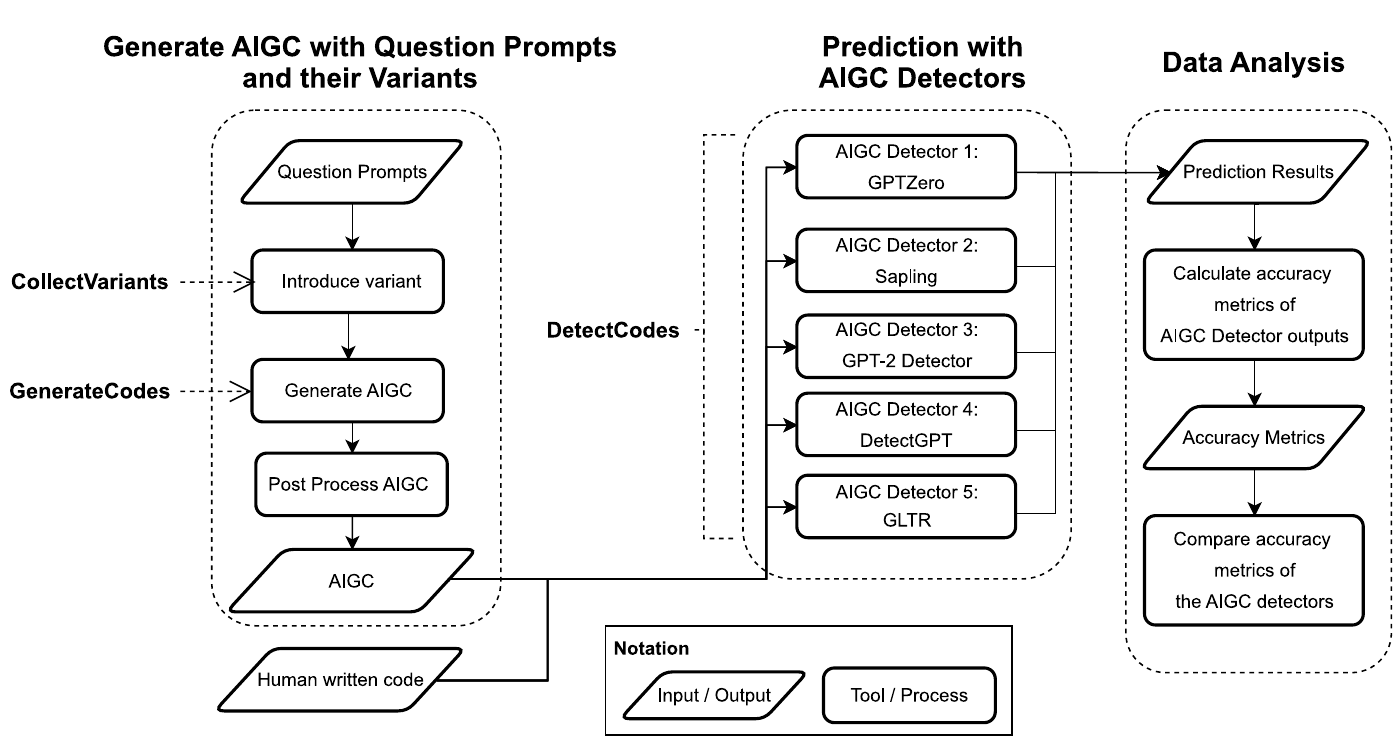}
  \setlength{\abovecaptionskip}{1em}
  \setlength{\belowcaptionskip}{-2em}
  \caption{Workflow of the AIGC and their variants' generation process, AIGC prediction with AIGC Detectors, and comparative analysis of AIGC Detector outputs with accuracy metrics.}
  \label{fig:Methodology}
  \vspace{1em}
\end{figure*}
\section{Background and Motivations}
In this section, we discuss some of the background related to our work, particularly in the domain of software engineering education and the use of generative AI technologies to help in teaching and learning processes. 

\subsection{Impact of AIGC on SE and CS Education}
Software Engineering (SE) and Computer Science (CS) education form the bedrock of technological progress \cite{zambonelli2002signs}. These fields equip students with essential skills, such as problem-solving, logical thinking, and creativity. However, the cornerstone of SE and CS education is programming. Programming is not just a skill; it embodies the essence of SE and CS education. It empowers students to create innovative software solutions, analyze complex problems, and make significant contributions to the digital world \cite{mirolo2022abstraction}.

The emergence of generative AI and the subsequent proliferation of AIGC have brought about a transformative era in education. AIGC represents a paradigm shift where automated content creation is revolutionizing learning experiences for both educators and learners \cite{grassini2023shaping}. This shift, while revolutionary, introduces a complex interplay of challenges and opportunities in the realm of education \cite{bozkurt2023generative}. AIGC can personalize and enhance learning, but it also poses questions about the authenticity of content and the methods of assessment in the digital age \cite{baidoo2023education}.

Generative Pre-trained Transformers (GPTs) \cite{ali2023gptcomparison}  have revolutionized natural language processing (NLP). It started with GPT-1 in 2018 (117 million parameters), laying the foundation. GPT-2 arrived in 2019 (1.5 billion parameters), improving text generation. GPT-3.5, part of OpenAI's API, introduced advanced features like editing and inserting. GPT-4 added multimodal capabilities, raising ethical concerns. Each iteration pushes AI language models forward with transformative capabilities.

The widespread adoption of AIGC has led to a growing need for reliable detection mechanisms. One notable development in this area is the AIGC Detector is detailed in the work by Wang et al. \cite{wang2023evaluating}. These detectors have demonstrated impressive accuracy in identifying AI-generated content. However, understanding the implications of the accuracy of AIGC Detector within educational settings, particularly in CS and SE learning environments, is essential.

In an era where these detectors ensure the authenticity of educational content, it is imperative to examine their effectiveness and limitations comprehensively. Relying solely on their accuracy might create a false sense of security, potentially allowing students to deceive these detectors. This raises questions about the robustness of educational assessments and the integrity of the learning process \cite{swiecki2022assessment}.

An intriguing aspect of this inquiry lies in potential scenarios where students might outsmart AIGC Detectors. For example, in specific situations, students may find ways to manipulate the system, deceiving even sophisticated AIs like ChatGPT \cite{Cemper_2023}. This raises concerns about the reliability of automated evaluations, especially if educators heavily depend on a single detector. Understanding these vulnerabilities is vital for ensuring the authenticity of educational assessments in the digital age \cite{krupiy2020vulnerability}.

In summary, this research is motivated by the transformative impact of AIGC on SE and CS education. Our study aims to significantly contribute by unraveling the complexities surrounding AIGC Detectors, exploring their accuracy, vulnerabilities, and implications for educational practices. By shedding light on these critical aspects, our work strives to enhance the understanding of AIGC Detection within the context of CS and SE education. Ultimately, our research seeks to ensure the integrity of educational assessments, thereby fostering a secure and genuine learning environment for students and educators alike.

\subsection{Selection of AIGC Detectors}
The integration of AIGC Detection in SE education significantly enhances academic integrity and ensures the authenticity of educational content. Educators can leverage these detectors to identify instances of plagiarism, unauthorized use of AI-generated code, and academic dishonesty among students, thereby fostering fairness and promoting originality within SE coursework. Moreover, AIGC Detectors play a pivotal role in verifying the authenticity of educational materials, allowing educators to maintain the credibility of resources utilized in SE courses. This technological integration not only strengthens the overall educational ecosystem but also stimulates meaningful discussions among students about responsible AI usage, emphasizing the importance of ethical practices in SE education.

Five AIGC Detectors are instrumental in achieving these objectives. GPTZero \cite{gptzero}, Sapling \cite{sapling}, GPT-2 Detector \cite{thien2023gpt2outputdataset}, DetectGPT \cite{detectgptRepository, mitchell2023detectgpt}, and Giant Language Model Test
Room (GLTR) \cite{guo-etal-2023-hc3, strobelt2023gltr, DBLP:journals/corr/abs-1906-04043} offer educators reliable tools to preserve academic integrity and encourage ethical AI usage. However, it is crucial for educators to thoroughly evaluate these tools and consider their limitations to ensure their effective use in SE education.

\section{Empirical Study Design and Methodology}
This section discusses the research questions, methodology and process of our empirical study. Figure \ref{fig:Methodology} offers an overview of the empirical study process, highlighting the key phase of data collection, prediction, and analysis. We have made the replication package publicly available.\footnote{\label{codedata}https://figshare.com/articles/dataset/Replication\textunderscore{}Package/24298036}

The data collection phase is essential for ensuring the integrity and robustness of our study. We gathered coding problems from various sources and introduced different variants based on the collected coding problems. Then, by utilizing OpenAI API service,\footnote{\label{openaiapi}https://platform.openai.com/} we collected AI-generated code and performed manual validation to ensure the correctness of the dataset. Moving on to the data prediction phase, by using AI-generated code and human-written code as input, we employed selected AIGC Detectors to perform AIGC Detection and collect the prediction results. In the data analysis phase, using the compiled prediction results, we evaluated the efficacy of each AIGC Detector via predetermined performance metrics and utilised the results analysed to answer our research questions.

\subsection{Research Questions and Experimental Setup\label{sec:experimentsetup}}
In Section \ref{sec:experimentsetup}, we discuss the research questions and the experimental setup.
\begin{enumerate}[leftmargin=*]
    \item \textbf{RQ1: How accurate are existing AIGC Detectors at detecting AI-generated code?}
    \item \textbf{RQ2: What are the limitations of existing AIGC Detectors when it comes to detecting AI-generated code?}
\end{enumerate}
Our objectives of RQ1 and RQ2 is to assess the performance of the chosen AIGC Detectors in identifying AI-generated code and investigate the limitations of AIGC Detectors in identifying AI-generated code. Both research questions share a similar experimental setup.
\begin{enumerate}[leftmargin=*]
  \item \textbf{Data Evaluation:} Evaluation of the AIGC Detectors' performance is based on the 13 variants of the dataset we have meticulously collected.
  \item \textbf{Experimental Configuration:} For the selected AIGC Detectors, we have established a discriminative threshold of 0.5. This threshold serves as a criterion to distinguish AI-generated code samples. Specifically, if the output probability surpasses the 0.5 threshold, it signifies the presence of AI-generated content within the input samples.
  \item \textbf{Performance Assessment Metrics:} The performance of the detectors is measured using a range of metrics such as accuracy, precision, true positive rate (TPR), false positive rate (FPR), true negative rate (TNR) and false negative rate (FNR). These metrics provide a comprehensive understanding of how effectively the detectors identify AI-generated code.
\end{enumerate}

\subsection{Data Collection}
We gathered fundamental code problems along with corresponding solutions code from various online sources, such as using the existing datasets from Kaggle \cite{kaggledescription} and web scraping technique to crawl data from Quescol \cite{quescoldescription}. Kaggle is an online community of data scientists and machine learning engineers that allows users to find datasets used in building AI models, publish datasets, collaborate with others, and enter competitions to solve data science challenges. Quescol is an educational website that provides a huge collection of the previous year and most important questions and answers as per the exam perspective. For the AIGC dataset, we generated different variants of AI-generated code to explore different scenarios and find out situations that might be possible for students to fool AIGC Detectors. The details about the different variants are discussed in Section \ref{sec:datavariation}.

For the compilation of the AIGC dataset, we developed an automation script leveraging the API service from OpenAI.\footref{openaiapi} The script will use the collected code problems as inputs, and then prompt ChatGPT to generate the outputs, subsequently label and store them as AI-generated code. The script was executed 13 times to generate different code variants, such as removing stop words or adding dead code, ensuring that a dataset of initial size 70,966 was collected for thorough analysis.

The coding problems and human-written code are obtained from three primary sources: 1.) {\verb|Python Coding Question|} \cite{quescolpython}: this source from Quescol contributed 80 samples to our dataset, which includes some common interview questions for Python; 2.) {\verb|Coding Problems and Solution Python Code|} \cite{kagglenaturallang}: this source from Kaggle contributes approximately 3,264 samples to our dataset, which encompasses a wide range of Python code solutions to various coding problems; 3.) {\verb|LeetCode Solutions and Content KPIs|} \cite{kaggleleetcodesolution}: this source from Kaggle contributes around 1,725 samples to our dataset which includes solutions to coding problems from the popular online coding platform, LeetCode, along with additional content and key performance indicators.
The collected dataset comprises 5,069 samples where each sample consists of a textual description of a coding problem and the corresponding solution code that were labeled as human-written code. The dataset covers diverse programming concepts, algorithms, and coding challenges. Based on collected code problems, we have created 13 variants of AIGC content.

{
\begin{algorithm}
\footnotesize
\caption{\textbf{DynCodeMetrics: } Dynamic Prompt Variation and Code Detection with Metric Calculation}
\begin{description}
\item[\textbf{Input:}]
\end{description}
\begin{itemize}
  \item $P$: Collection of Prompt Dataset
  \item $AIGCD$: Artificial Intelligence Generated Content Detector
  \item $VarPrompt$: Variant Modification
\end{itemize}

\begin{description}
\item[\textbf{Output:}] $R$: Metric Results
\end{description}

\begin{algorithmic}[1]
  \Procedure{DynCodeMetrics}{$P$, $AIGCD$, $VarPrompt$}
  \State $V \gets \{\} $ \text{  // Collection of various prompts}
  \State $D \gets \{\}$ \text{  // Collection Results of 1(Human) and 0(AI)}
  \State $C \gets \{\}$ \text{  // Collection of Human and AI-generated Code}
  
  \State $V \gets$ \Call{CollectVariants}{$P$, $VarPrompt$, $V$}
  \State \Call{GenerateCodes}{$V$, $C$}
  \State $C' \gets$ Post Process on $C$ \text{  // Post-processing on generated codes}
  \State \Call{DetectCodes}{$AIGCD$, $C'$, $D$}
  \State $R \gets$ Calculate accuracy, precision, recall, TPR, FPR based on $D$
  
  \State \textbf{return} $R$ \text{  // Result of Metrics}
  \EndProcedure
\end{algorithmic}
\end{algorithm}

\begin{algorithm}
\footnotesize
\caption{\textbf{CollectVariants: } Generate Different Variation of Prompt}
\begin{description}
\item[\textbf{Input:}]
\end{description}
\begin{itemize}
  \item $P$: Collection of Prompt Dataset
  \item $VarPrompt$: Variant Modification
  \item $V$: Collection of various prompts
\end{itemize}

\begin{description}
\item[\textbf{Output:}] $V$: Collection of Various Prompts
\end{description}

\begin{algorithmic}[1]
  \Procedure{CollectVariants}{$P$, $VarPrompt$, $V$}
  \For{$p \in P$}
    \State $v \gets$ Variant Modification on $p$
    \State Include $v$ in collection of various prompts $V$
  \EndFor
  \State \textbf{return} $V$ \text{  // Collection of various prompts}
  \EndProcedure
\end{algorithmic}
\end{algorithm}

\begin{algorithm}
\footnotesize
\caption{\textbf{GenerateCodes: } Collecting AI-Generated Code}
\begin{description}
\item[\textbf{Input:}]
\end{description}
\begin{itemize}
  \item $V$: Collection of various prompts
  \item $C$: Data Collection of Human and AI-generated Code
\end{itemize}

\begin{description}
\item[\textbf{Output:}] $C$: Modified Data Collection
\end{description}

\begin{algorithmic}[1]
  \Procedure{GenerateCodes}{$V$, $C$}
  \For{$v \in V$}
    \State $C \gets \{\}$ \text{  // Clear the Data Collection}
    \For{$i \gets 1$ to $N$} \text{  // Generate AI Code}
      \State $c \gets$ AI Code Generation based on $v$
      \State Include $c$ in Data Collection $C$
    \EndFor
  \EndFor
    \State \textbf{return} $C$ \text{  // Code Data Collection}
  \EndProcedure
\end{algorithmic}
\end{algorithm}

\begin{algorithm}
\footnotesize
\caption{\textbf{DetectCodes: } Code Detection Using AIGC Detector}
\begin{description}
\item[\textbf{Input:}]
\end{description}
\begin{itemize}
  \item $AIGCD$: Artificial Intelligence Generated Content Detector
  \item $C'$: Processed Data Collection of Human and AI-generated Code
  \item $D$: Collection Results of 1(Human) and 0(AI)
\end{itemize}

\begin{description}
\item[\textbf{Output:}] $D$: Detection Result Collection
\end{description}

\begin{algorithmic}[1]
  \Procedure{DetectCodes}{$AIGCD$, $C'$, $D$}
  \For{$aigcd \in AIGCD$}
    \State $D \gets \{\}$ \text{  // Result Collection of 1 (Human) and 0 (AI)}
    \For{$c \in C'$}
      \State $d \gets$ Code Detection using $AIGCD$
      \State Include $d$ in Result Collection $D$
    \EndFor
  \EndFor
  \State \textbf{return} $D$ \text{  // Detection Result Collection}
  \EndProcedure
\end{algorithmic}
\end{algorithm}
}

\subsection{Workflow Explanation}
To ensure the replicability and clarity of our empirical study procedure, we have provided our procedure in a systematic, step-by-step algorithm format to eliminate misinterpretation of our procedure in Figure \ref{fig:Methodology}.
As shown in Algorithms 1 to 4, four procedures are included on how variant dataset is collected and metric is calculated:

\begin{itemize}[leftmargin=*]
    \item{\verb|DynCodeMetrics|}: Algorithm 1, named \textbf{DynCodeMetrics}, plays a central role in coordinating various sub-algorithms to streamline the workflow of dynamic prompt generation, AI-generated code post-processing, code detection, and metric computation. Taking inputs $P$ (Prompt Datasets), $AIGCD$ (AIGC Detector), and VarPrompt (Variant Modification Method), it initializes collections $V$ (Variant Prompts), $D$ (AIGC Detector Results), and $C$ (Human and AI-generated Codes). The procedure first applies the \textbf{CollectVariants} method to modify prompts using variants and stores them in $V$. Subsequently, it employs \textbf{GenerateCodes} to create AI-generated codes based on modified prompts and performs post-processing on $C$ to clean and validate the dataset. The \textbf{DetectCodes} method utilizes the AIGC Detector to identify human and AI-generated codes, updating $D$ accordingly. The algorithm concludes by calculating accuracy, precision, true positive rate (TPR), false positive rate (FPR), true negative rate (TNR), and false negative rate (FNR) based on detection results, providing a comprehensive analysis. The calculated metrics, denoted as $R$, are returned as the output.
    \item{\verb|CollectVariants|}: From Algorithm 2, the \textbf{CollectVariants} takes three inputs: $P$ (Collection of Prompts Datasets), $VarPrompt$ (Variant Modification Method), and empty collection $V$ (collection of various prompts). It iterates over each coding problem in the dataset and applies the variant modification method to modify the prompt. The modified prompts are then included in the $V$ (collection of various prompts) and being returned.
    \item{\verb|GenerateCodes|}: From Algorithm 3, the \textbf{GenerateCodes} takes $V$ (collection of prompts) and empty collection $C$ (collection of Human and AI-generated codes) as input. It iterates each variant prompt and generates codes based on the variant prompt. The procedure generates the specified number of AI codes for each variant prompt and includes them in the data collection. The data collection $C$ (Collection of Human and AI-generated codes) contains the code from humans and AI are returned.
    \item{\verb|DetectCodes|}: From Algorithm 4, the \textbf{DetectCodes} procedure takes the $AIGCD$ (AIGC Detector), $C'$ (Postprocessed Human and AI-generated codes), and an empty $D$ collection (collection of AIGC Detector's results) for storing the detection results as input. It iterates over each AIGC Detector and performs classification on each code in the post-processed data using the corresponding model. The detection results are then included in the result $D$ collection (collection of AIGC Detector's results).
\end{itemize}
Algorithm 1 combines Algorithms 2 to 4 to dynamically vary prompts, generate AI codes, perform code detection, and calculate relevant metrics. The resulting metrics can be used to evaluate the effectiveness and accuracy of the AIGC Detector.
{
\tiny
\begin{table*}[htbp]
\centering
\caption{Post-processed dataset size for each Detector across variant}
\footnotesize
\begin{tabular}{l *{13}{c}}
\toprule
\multirow{2}{*}{\textbf{Detector}} & \multicolumn{13}{c}{\textbf{Variant}} \\
\cmidrule(lr){2-14}
 & 1 & 2 & 3 & 4 & 5 & 6 & 7 & 8 & 9 & 10 & 11 & 12 & 13 \\
\midrule
GLTR           & 5,069 & 5,065* & 5,064* & 5,069 & 5,069 & 5,069 & 5,069 & 5,069 & 5,069 & 5,069 & 5,069 & 5,069 & 5,068* \\
Sapling        & 5,069 & 5,055* & 5,064* & 5,069 & 5,069 & 5,069 & 5,069 & 5,069 & 5,069 & 5,069 & 5,069 & 5,069 & 5,068*  \\
GPT Zero       & 5,069 & 5,065* & 5,064* & 5,069 & 5,069 & 5,069 & 5,069 & 5,069 & 5,069 & 5,069 & 5,069 & 5,069 & 5,068* \\
GPT-2 Detector & 5,069 & 5,065* & 5,064* & 5,069 & 5,069 & 5,069 & 5,069 & 5,069 & 5,069 & 5,069 & 5,069 & 5,069 & 5,068* \\
DetectGPT      & 4,016* & 4,674* & 4,509* & 4,097* & 4,785* & 4,702* & 4,888* & 3,932* & 3,997* & 3,952* & 4,211* & 3,925* & 4,785*  \\
\bottomrule
\end{tabular}
\textit{\\Note: * indicates a deviation from the initial dataset size of 5,069.}

\label{tab:variant_size}
\end{table*}
}

\subsection{Data Variations\label{sec:datavariation}}
Table \ref{tab:variant_size} displays the dataset sizes used for each AIGC Detector across various variants. The sizes of datasets for Variants 2 and 3 (we will discuss the specific details of each variant in the following paragraph) are smaller across all AIGC Detectors due to the removal of data resulting from invalid outputs by ChatGPT. Specifically, when applying the Sapling detector on Variant 2, the dataset size is further reduced as our initial approach has a minimum word requirement for code detection (which will be discussed in detail in the following paragraph). Similarly, the DetectGPT detector excludes data with fewer than 100 words due to its minimum word requirement, leading to a reduced dataset size.

In the process of preparing our dataset, we introduced 13 variations of AI-generated code. These variations were achieved by altering the prompts provided to ChatGPT or the solution code received from ChatGPT. Our aim was to find out the limitations and performance of each AIGC Detector under different variants of programming code. Detailed descriptions of each variation are available on Figshare. \footnote{https://figshare.com/articles/dataset/Variant\textunderscore{}Description/24265018}
{
\vspace{0.2pt}
\\
\noindent \textbf{1. Without modification}: Adapted the prompt based on Sam Altman's example\footnote{\label{openaiceotweet}https://twitter.com/sama/status/1682826943312326659}. Prompted ChatGPT with unmodified code problems, adding preliminary conditions\cite{chatgptPromptEngineeringGuidelines}. Mimics users' typical approach to resolving coding problems.
\\
\noindent \textbf{2. Removal of Stopwords}: Preprocessed coding problems by removing common stopwords using the NLTK library. Aims to simulate users modifying questions to enhance output by eliminating stopwords.
\\
\noindent \textbf{3. Ask to Mimic Human}: Appended "Please mimic a human response." to each prompt, guiding ChatGPT to generate contextually appropriate code mimicking human responses and enhancing its ability to mimic human-written code.
\\
\noindent \textbf{4. Solution Without Comment}: Instructed ChatGPT to exclude comments in the solution, simulating cases where students omit comments, violating good programming practices.
\\
\noindent \textbf{5. Assertion Test Code}: Instructed ChatGPT to include assertion test code, assessing AIGC Detectors' ability to identify syntactically valid code with test assertions—a common approach for verifying program correctness.
\\
\noindent \textbf{6. Solution with Test Case}: Instructed ChatGPT to include test cases, evaluating AIGC Detectors' ability to identify syntactically valid code with included test cases—a common method for verifying program correctness.
\\
\noindent \textbf{7. Unittest Test Case}: Instructed ChatGPT to include unittest test cases, evaluating AIGC Detectors' ability to identify syntactically valid code with included unittest test cases—a common approach for verifying program correctness.
\\
\noindent \textbf{8. Replace variable names}: Using the AI-generated code from variant 1, we used the AST library to replace all variable names with single-character letters (a to z). Simulates cases where students violate basic programming rules in variable naming.
\\
\noindent \textbf{9. Replace function names}: Using the AI-generated code from variant 1, we developed a Python script with the AST library to replace all function names with single-character letters (a to z). Simulates cases where students violate basic programming rules in function naming.
\\
\noindent \textbf{10. Replace variable and function names}: Using AI-generated code from variant 1, we developed a Python script with the AST library to replace all variable and function names with single-character letters (a to z). Simulates extreme cases where programmers violate basic programming rules in naming convention.
\\
\noindent \textbf{11. Long method}: Long method, a term associated with code smells, indicates design weaknesses that raise the risk of bugs. Asked ChatGPT for a longer output in function block format to simulate scenarios with long method code smell, reducing code readability and maintainability.
\\
\noindent \textbf{12. Short method}: Contrary to the long method variant, prompted ChatGPT to generate a shorter output in function block format. Aims to simulate scenarios adhering to best programming practices with concise methods focusing on a single functionality.
\\
\noindent \textbf{13. Adding 5 snippets of dead code}: Injected 5 snippets of dead code into the output as noise for the AIGC Detector. Aims to explore the impact of dead code on the detector's performance, simulating scenarios where useless code is inadvertently included in the solution during development.
}

\subsection{Evaluation Metrics}
\begin{table}[]
\footnotesize
\centering
\caption{Confusion Matrix used in this study}
\begin{tblr}{
  cells = {c},
  cell{1}{1} = {r=4}{},
  cell{1}{3} = {c=2}{},
  vline{2} = {3-4}{},
  vline{3-5} = {2-4}{},
  hline{2} = {3-4}{},
  hline{3-5} = {2-4}{},
}
\begin{sideways}Predicted Value\end{sideways} &  & Actual Value        &                     \\
&           & HUMAN (1)           & AI (0)              \\
& HUMAN (1) & True Positive (TP)  & False Positive (FP) \\
& AI (0)    & False Negative (FN) & True Negative (TN)  
\end{tblr}
\label{tab:metrics_base}
\end{table}

For this empirical study, we have selected a few metrics~\cite{kulkarni2020foundations, vujovic2021classification} to evaluate the performance of selected AIGC Detectors. Table \ref{tab:metrics_base} shows the confusion matrix used in the paper where 1 indicates human-written code, and 0 indicates AI-generated code.

\textbf{TPR/Recall:} True Positive Rate, also known as Recall, calculated as $TPR/Recall=\frac{TP}{TP+FN}$,
where TP is the number of human-written codes correctly labelled as human-written, FN is the number of human-written codes incorrectly labelled as AI-generated and TP+FN represents the total number of human-written codes.

\textbf{FNR:} False Negative Rate, calculated as $FNR=\frac{FN}{TP+FN}$,
where FN is the number of human-written codes incorrectly labelled as AI-generated, TP is the number of human-written codes correctly labelled as human-written and TP+FN represents the total number of human-written codes.

\textbf{TNR:} True Negative Rate, calculated as $TNR=\frac{TN}{TN+FP}$,
where TN is the number of AI-generated codes correctly labelled as AI-generated, FP is the number of AI-generated codes incorrectly labelled as human-written and TN+FP represents the total number of AI-generated codes.

\textbf{FPR:} False Positive Rate, calculated as $FPR=\frac{FP}{TN+FP}$,
where FP is the number of AI-generated codes incorrectly labelled as human-written, TN is the number of AI-generated codes correctly labelled as AI-generated and TN+FP represents the total number of AI-generated codes.

\textbf{Accuracy (ACC):} Calculated as 
$Accuracy=\frac{TP+TN}{TP+TN+FP+FN}$,
where TP is the number of human-written codes correctly labelled as human-written, TN is the number of AI-generated codes correctly labelled as AI-generated and TP+TN+FP+FN is the total number of codes.

\textbf{Precision:} Calculated as $Precision=\frac{TP}{TP+FP}$,
where TP is the number of human-written codes correctly labelled as human-written, FP is the number of AI-generated codes incorrectly labelled as human-written and TP+FP is the total number of codes labelled as human-written.

\textbf{F1 Score:} Calculated as $F_{1}=\frac{2}{\frac{1}{Recall}+\frac{1}{Precision}}$,
which represents the harmonic mean of precision and recall.

\textbf{AUC:} Area Under Curve, which is used to evaluate how much the model is capable of distinguishing between classes. A higher AUC score means that it has a better capability to distinguish between positive and negative classes. When the AUC score is around 0.5, it indicates the model performs random choice when predicting the samples.

\section{Results}
In this section, we showcase our experimental results and provide an analysis to address each research question. We have uploaded the full set of results on Figshare.\footnote{\label{Result Table}https://figshare.com/articles/dataset/Variant\textunderscore{}Result/24265015}

\subsection{RQ1: How accurate are existing AIGC Detectors at detecting AI-generated code?}

\begin{table*}[htbp]
\centering
\caption{Accuracy, TPR, and TNR of All 5 AIGC Detectors}
\label{tab:detector-results}
\footnotesize 
\begin{tabular}{ll*{13}{c}}
\toprule
\multirow{2}{*}{\textbf{Detector}} & \multirow{2}{*}{\textbf{Metric}} & \multicolumn{13}{c}{\textbf{Variant}} \\
\cmidrule(lr){3-15}
 & & 1 & 2 & 3 & 4 & 5 & 6 & 7 & 8 & 9 & 10 & 11 & 12 & 13 \\
\midrule

\multirowcell{3}[0pt][l]{GPT Zero} & ACC & 0.4971 & \textcolor{lgreen}{ 0.4988} & \textcolor{lgreen}{ 0.4986} & \textcolor{red}{ 0.4966} & \textcolor{red}{ 0.4969} & \textcolor{lgreen}{ 0.4976} & \textcolor{lgreen}{ 0.4989} & \textcolor{lgreen}{ 0.5108} & \textcolor{lgreen}{ 0.4972} & \textcolor{lgreen}{ 0.4973} & \textcolor{red}{ 0.4970} & \textcolor{red}{ 0.4965} & \textcolor{lgreen}{ 0.5824}  \\ 
 & TPR & 0.9927 & 0.9927 & 0.9927 & 0.9927 & 0.9927 & 0.9927 & 0.9927 & 0.9927 & 0.9927 & 0.9927 & 0.9927 & 0.9927 & 0.9927 \\ 
 & TNR & 0.0016 & \textcolor{lgreen}{ 0.0049} & \textcolor{lgreen}{ 0.0045} & \textcolor{red}{ 0.0006} & \textcolor{red}{ 0.0012} & \textcolor{lgreen}{ 0.0026} & \textcolor{lgreen}{ 0.0051} & \textcolor{lgreen}{ 0.0288} & \textcolor{lgreen}{ 0.0018} & \textcolor{lgreen}{ 0.0020} & \textcolor{red}{ 0.0014} & \textcolor{red}{ 0.0004} & \textcolor{lgreen}{ 0.1721} \\ 

\cmidrule(lr){2-15}
\multirowcell{3}[0pt][l]{GPT-2 Detector} & ACC & 0.5043 & \textcolor{red}{ 0.5005} & \textcolor{red}{ 0.4961} & \textcolor{red}{ 0.5013} & \textcolor{red}{ 0.4716} & \textcolor{red}{ 0.4788} & \textcolor{red}{ 0.4828} & \textcolor{lgreen}{ 0.5134} & \textcolor{lgreen}{ 0.5100} & \textcolor{lgreen}{ 0.5252} & \textcolor{red}{ 0.4911} & \textcolor{red}{ 0.4958} & \textcolor{red}{ 0.4922}   \\ 
 & TPR & 0.9128 & 0.9127 & 0.9127 & 0.9128 & 0.9128 & 0.9128 & 0.9128 & 0.9128 & 0.9128 & 0.9128 & 0.9128 & 0.9128 & 0.9128 \\ 
 & TNR & 0.0959 & \textcolor{red}{ 0.0883} & \textcolor{red}{ 0.0796} & \textcolor{red}{ 0.0898} & \textcolor{red}{ 0.0304} & \textcolor{red}{ 0.0448} & \textcolor{red}{ 0.0529} & \textcolor{lgreen}{ 0.1140} & \textcolor{lgreen}{ 0.1071} & \textcolor{lgreen}{ 0.1375} & \textcolor{red}{ 0.0694} & \textcolor{red}{ 0.0787} & \textcolor{red}{ 0.0716} \\ 

\cmidrule(lr){2-15}
\multirowcell{3}[0pt][l]{DetectGPT} & ACC & 0.4893 & \textcolor{red}{ 0.4742} & \textcolor{red}{ 0.4685} & \textcolor{lgreen}{ 0.4941} & \textcolor{red}{ 0.4055} & \textcolor{red}{ 0.4014} & \textcolor{lgreen}{ 0.5132} & \textcolor{lgreen}{ 0.4943} & \textcolor{lgreen}{ 0.5278} & \textcolor{lgreen}{ 0.5125} & \textcolor{lgreen}{ 0.5354} & \textcolor{lgreen}{ 0.5153} & \textcolor{red}{ 0.4373} \\ 
 & TPR & 0.2756 & 0.2993 & 0.2978 & 0.2824 & 0.3083 & 0.3056 & 0.3124 & 0.2693 & 0.2737 & 0.2677 & 0.2769 & 0.2662 & 0.3093 \\ 
 & TNR & 0.7029 & \textcolor{red}{ 0.6491} & \textcolor{red}{ 0.6392} & \textcolor{lgreen}{ 0.7059} & \textcolor{red}{ 0.5028} & \textcolor{lgreen}{ 0.4972} & \textcolor{lgreen}{ 0.7140} & \textcolor{lgreen}{ 0.7192} & \textcolor{lgreen}{ 0.7818} & \textcolor{lgreen}{ 0.7573} & \textcolor{lgreen}{ 0.7939} & \textcolor{lgreen}{ 0.7643} & \textcolor{red}{ 0.5653} \\ 

\cmidrule(lr){2-15}
\multirowcell{3}[0pt][l]{GLTR} & ACC & 0.5040 & \textcolor{red}{ 0.4936} & \textcolor{red}{ 0.4841} & \textcolor{lgreen}{ 0.6569} & \textcolor{lgreen}{ 0.6999} & \textcolor{lgreen}{ 0.6920} & \textcolor{lgreen}{ 0.7693} & \textcolor{red}{ 0.4908} & \textcolor{red}{ 0.4952} & \textcolor{red}{ 0.4881} & \textcolor{lgreen}{ 0.5375} & \textcolor{red}{ 0.5020} & \textcolor{lgreen}{ 0.6478} \\ 
 & TPR & 0.6461 & 0.6464 & 0.6461 & 0.6461 & 0.6461 & 0.6461 & 0.6461 & 0.6461 & 0.6461 & 0.6461 & 0.6461 & 0.6461 & 0.6460 \\ 
 & TNR & 0.3620 & \textcolor{red}{ 0.3408} & \textcolor{red}{ 0.3221} & \textcolor{lgreen}{ 0.6678} & \textcolor{lgreen}{ 0.7538} & \textcolor{lgreen}{ 0.7378} & \textcolor{lgreen}{ 0.8925} & \textcolor{red}{ 0.3356} & \textcolor{red}{ 0.3442} & \textcolor{red}{ 0.3300} & \textcolor{lgreen}{ 0.4289} & \textcolor{red}{ 0.3579} & \textcolor{lgreen}{ 0.6496} \\ 

\cmidrule(lr){2-15}
\multirowcell{3}[0pt][l]{Sapling} & ACC & 0.6056 & \textcolor{red}{ 0.5961} & \textcolor{red}{ 0.6031} & \textcolor{red}{ 0.6048} & \textcolor{red}{ 0.5425} & \textcolor{red}{ 0.5828} & \textcolor{lgreen}{ 0.6083} & \textcolor{lgreen}{ 0.6630} & \textcolor{lgreen}{ 0.6258} & \textcolor{lgreen}{ 0.6811} & \textcolor{lgreen}{ 0.6187} & \textcolor{lgreen}{ 0.6059} & \textcolor{lgreen}{ 0.6528} \\ 
 & TPR & 0.4780 & 0.4797 & 0.4783 & 0.4780 & 0.4780 & 0.4780 & 0.4780 & 0.4780 & 0.4780 & 0.4780 & 0.4780 & 0.4780 & 0.4781 \\ 
 & TNR & 0.7333 & \textcolor{red}{ 0.7126} & \textcolor{red}{ 0.7279} & \textcolor{red}{ 0.7315} & \textcolor{red}{ 0.6070} & \textcolor{red}{ 0.6875} & \textcolor{lgreen}{ 0.7386} & \textcolor{lgreen}{ 0.8479} & \textcolor{lgreen}{ 0.7735} & \textcolor{lgreen}{ 0.8842} & \textcolor{lgreen}{ 0.7593} & \textcolor{lgreen}{ 0.7339} & \textcolor{lgreen}{ 0.8275} \\ 
\bottomrule
\end{tabular}
\textit{\\Note: \textcolor{lgreen}{green} indicates an increase compared to Variant 1, while \textcolor{red}{red} indicates a decrease}
\label{tab:TPR_TNR_ACC}
\end{table*}

Table \ref{tab:TPR_TNR_ACC} shows the performance of GPT-2 Detector and GPTZero, where both detectors exhibit poor performance when compared with baseline variant 1. Their Accuracy (ACC) results hover around 0.5, suggesting a lack of effectiveness in distinguishing AI-generated code from human-written code. They tend to classify a significant portion of input code as human-generated rather than machine-generated. This performance issue could stem from their primary training on natural language data or potential overfitting to natural language.

In the assessment of AIGC Detectors, both the GPT-2 Detector and GPTZero exhibited challenges in effectively distinguishing between human-written and AI-generated code across all evaluated variants. In Table \ref{tab:TPR_TNR_ACC}, their performance was characterized by high TPR and low TNR while having an ACC of around 0.5, indicating a tendency to classify both human-written and AI-generated code as human-written code.

\begin{figure}[t]
  \centering
  \tiny
  \includegraphics[scale=0.28]{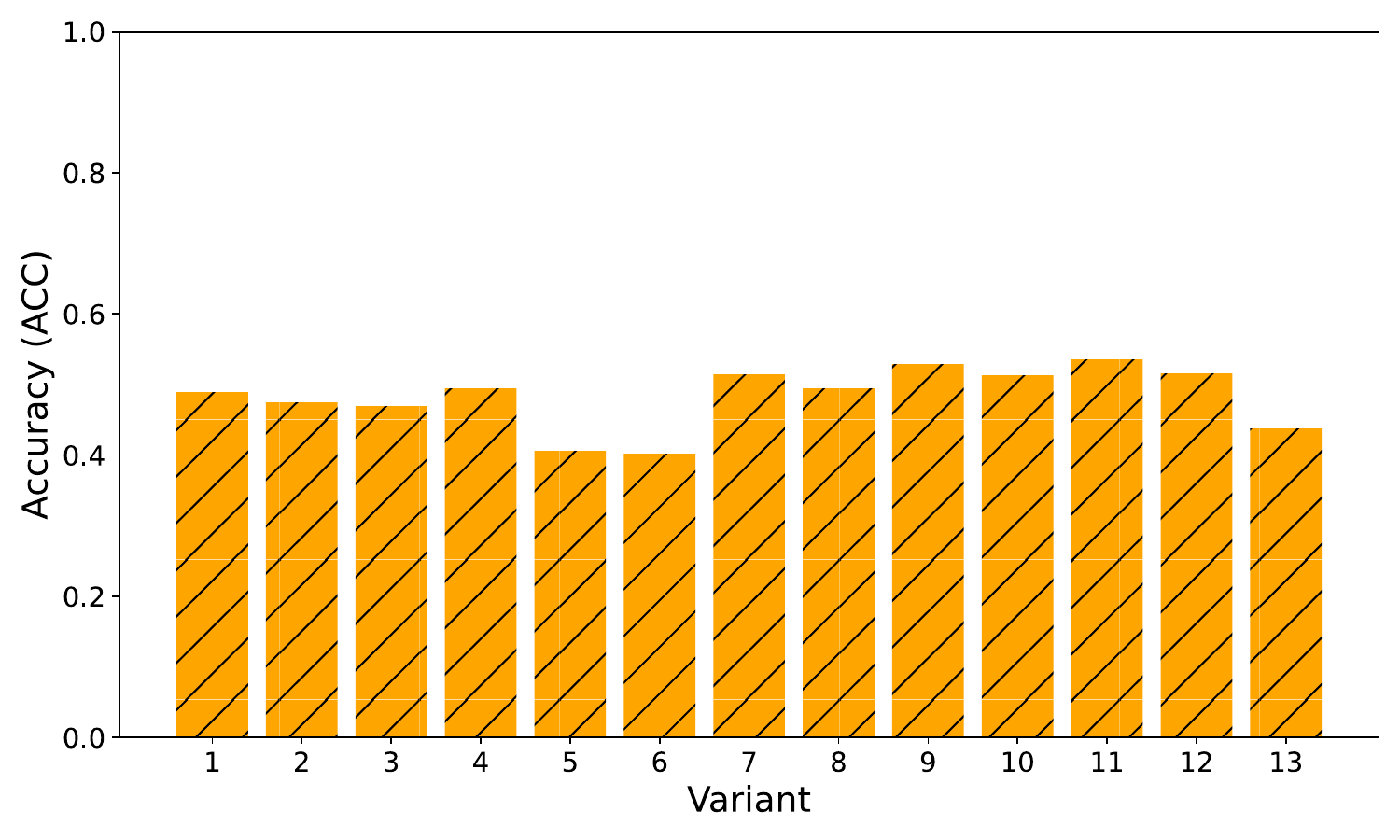}
  \setlength{\abovecaptionskip}{1em}
  \setlength{\belowcaptionskip}{-1em}
  \caption{Accuracy Performance for DetectGPT}
  \label{fig:ACC DetectGPT}
  \vspace{-2em}
\end{figure}
DetectGPT, akin to the aforementioned AIGC Detectors, showcases an ACC near 0.5 in Figure \ref{fig:ACC DetectGPT}, indicating its struggle in distinguishing AI-generated from human-written code. Notably, DetectGPT tends to misclassify the majority of human-generated code as AI-generated.
However, a more positive aspect emerges when we compare DetectGPT to the baseline (Variant 1) and other variants like Function Name (Variant 9), Variable and Function Name (Variant 10), Long Method (Variant 11), and Short Method (Variant 12). DetectGPT demonstrates an improvement in TNR ranging from 5\% to 9\%. This improvement signifies its competence in accurately identifying AIGC.

\begin{figure}[t]
  \centering
  \tiny
  \includegraphics[scale=0.28]{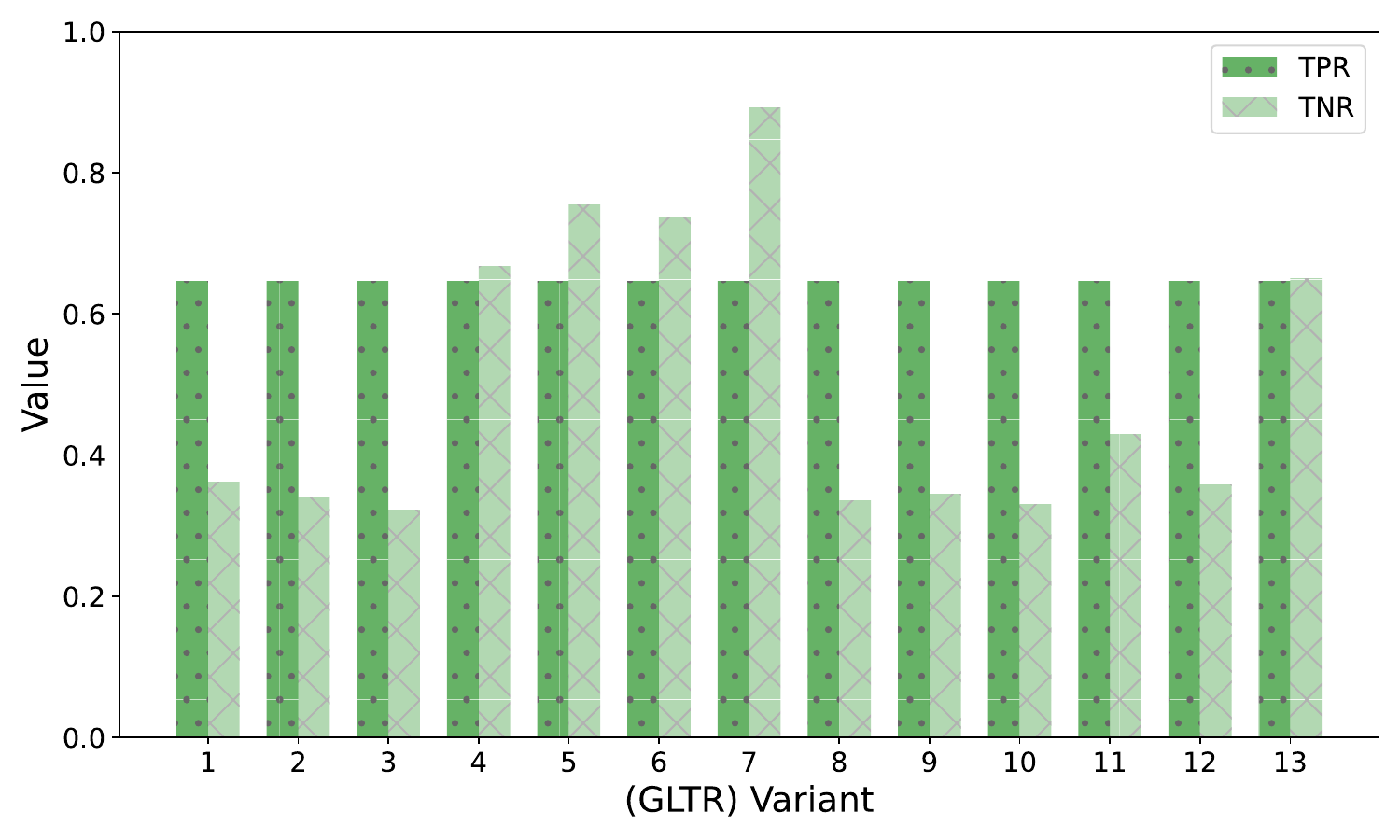}
  \setlength{\abovecaptionskip}{1em}
  \setlength{\belowcaptionskip}{-1em}
  \caption{TPR and TNR Performance for GLTR}
  \label{fig:TPR TNR GLTR}
  \vspace{-2em}
\end{figure}

The ACC performance of GLTR exhibits variability across different AIGC variants. Approximately half of the variants surpass the 0.6 Accuracy (ACC) threshold, while the remainder approach a more modest 0.5.
In our evaluation of GLTR and Figure \ref{fig:TPR TNR GLTR}, we conducted a comparative analysis involving the baseline Variant 1 and several other variants, including "No Comment (Variant 4)," "Assertion (Variant 5)," "Test Case (Variant 6)," "Unittest Test Case (Variant 7)," and "Dead Code (Variant 13)." The aim was to assess the system's performance in detecting GPT-generated code as AI-generated content.

The results uncovered a significant enhancement in GLTR's ability to identify AI-generated code when compared to the baseline variant. This improvement ranged impressively from 18\% to 53\% across the diverse set of variants. This suggests that the incorporation of these new variants, such as "No Comment", "Assertion", "Test Case", "Unittest Test Case", and "Dead Code", substantially bolstered the system's proficiency in AI-generated code detection.
Additionally, it is noteworthy that the GLTR exhibited a notable 6\% increase in the TNR when tasked with evaluating the "Long Method" variant 11. This improvement signifies an augmented capability to accurately categorize non-AI-generated code, particularly in the context of extensive and intricate code segments.

\begin{figure}[t]
  \centering
  \tiny
  \includegraphics[scale=0.28]{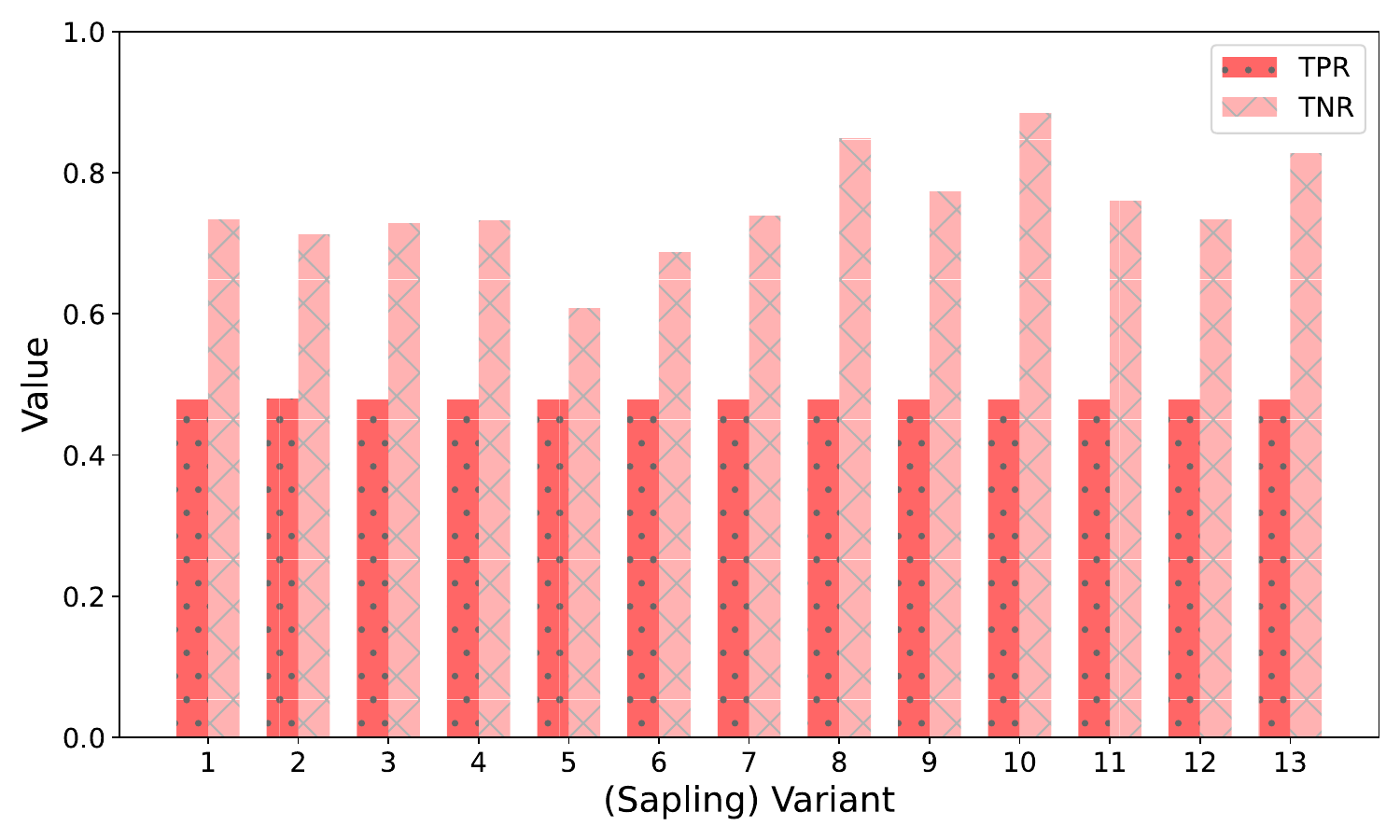}
  \setlength{\abovecaptionskip}{1em}
  \setlength{\belowcaptionskip}{-1em}
  \caption{TPR and TNR Performance for Sapling}
  \label{fig:TPR TNR Sapling}
  \vspace{-2em}
\end{figure}
Sapling detector outperforms other AIGC Detectors, consistently achieving ACC values above 0.6 in ten out of fourteen AIGC variants. In our evaluation of Sapling's AI Detector and Figure \ref{fig:TPR TNR Sapling}, we compared its performance to the baseline (Variant 1) across various code variants. Our findings highlight significant improvements, particularly in TNR, when the Sapling detector is applied to specific variants such as "Variable Name" (Variant 8), "Variable and Function Name" (Variant 10), and "DeadCode" (Variant 13). These variants exhibited a substantial increase in TNR, ranging from 9\% to 15\%, indicating their enhanced ability to detect AI-generated code accurately.

\begin{framed}
\small
\noindent{\textbf{Answer to RQ1:}} Existing AIGC Detectors perform poorly in distinguishing between human-written code and AI-generated code, indicating the inherent weaknesses of current detectors. This underscores the need for further research and development in this domain to enhance their efficacy.

\end{framed}

\subsection{RQ2: What are the limitations of existing AIGC Detectors when it comes to detecting AI-generated code?}
From the results in Table \ref{tab:detector-results}, we have revealed significant sensitivity to code variants, particularly with GLTR, as demonstrated by a Samples Paired t-Test \cite{chumney2018paired} comparing ACC differences between the original (Variant 1) and all other versions of variants (Variant 2 to 13), where the p-values for all five AIGC Detectors are 0.0296 (GLTR), 0.1032 (GPT-2 Detector), 0.2479 (GPTZero), 0.3816 (Sapling) and 0.5714 (DetectGPT). GLTR demonstrated the lowest p-value, signifying a significant performance difference across the variants. This underscores GLTR's heightened sensitivity to specific variant patterns, resulting in misclassifications, as evident in the wide range of accuracy from 0.4841 (Variant 3) to 0.7693 (Variant 7). This finding highlights a potential vulnerability of AIGC Detectors, particularly when students prompt ChatGPT to mimic a human's response when generating the code as GLTR tends to treat most AIGC incorrectly as human-written, potentially enabling academic plagiarism. Moreover, Variants 2, 3, and 5 induced a drop in performance (both ACC and TNR) across multiple models when compared to that of Variant 1, underlining the need for further research to address these specific vulnerabilities and ensure the integrity of AI-generated content detection methods in educational settings.

After some thorough analysis of each AIGC Detector's performance across each variant, we discuss some of the limitations that we observed for each AIGC Detector. 


{
\vspace{0.2pt}

\noindent \textbf{1. GPTZero Limitations}: GPTZero may struggle to accurately detect AI-generated code due to the distinct syntax and structure of programming languages. Code follows strict rules and conventions, deviating from the linguistic patterns GPTZero is designed to identify. This mismatch can result in limited effectiveness, leading to potential false positives or false negatives when applied to code detection tasks.
\\
\noindent \textbf{2. Sapling Limitations}: Sapling is designed for analyzing text generated by language models like GPT-3 and ChatGPT, excelling in identifying such content. However, it may not be optimized for detecting programming code, as AI-generated code can have distinct characteristics not aligned with natural language patterns. The limitations of Sapling AI Detector include its focus on language model-generated text and potential challenges in code detection.
\\
\noindent \textbf{3. GPT-2 Detector Limitations}: The GPT-2 Detector relies on a dataset of text generated by GPT-2 models, making it effective for identifying GPT-2 outputs. However, it may struggle with other AI models or programming code, as distinct characteristics might not be represented in the dataset. The detector's ability to detect code from different models is limited, given the variability in complexity and structure of programming code.
\\
\noindent \textbf{4. DetectGPT Limitations}: DetectGPT employs perturbations in input text for innovative text classification but may face limitations in detecting programming code. The intricate logic and syntax of code may not exhibit the same perturbation patterns as natural language, leading to challenges in adaptability and potential for false or missed detections in this context.
\\
\noindent \textbf{5. GLTR Limitations}: GLTR, relying on models like GPT-2, enhances text interpretability through word rankings and color coding. However, its limitations are apparent in programming code detection, where the unique linguistic patterns and nuances are not effectively captured by these methods. GLTR may struggle to provide accurate detections for programming code.
}

Our findings indicate a common limitation shared among current AIGC Detectors in their ability to detect AI-generated code: they consistently perform poorly. In our comprehensive assessment of various AI detection tools, we observed a lack of accuracy and reliability in their capacity to identify AI-generated code content.

These limitations encompass several key areas:
{
\vspace{0.2pt}
\\
\noindent \textbf{1. Detection Accuracy:} Across the board, these tools struggled to accurately identify AI-generated code, often producing high rates of false negatives and false positives, leading to inaccuracies.
\\
\noindent \textbf{2. Lack of Specificity:} Existing tools encountered difficulties distinguishing between human-generated and AI-generated code, resulting in misclassifications and a lack of precision.
\\
\noindent \textbf{3. Generalization Challenges:} While fine-tuning improved performance within specific code domains, these tools faced difficulty in generalizing their capabilities. They often faltered when confronted with code from diverse domains or when AI-generated code closely resembled human coding styles.
}

\begin{framed}
\small
\noindent{\textbf{Answer to RQ2:}} The limitations of AIGC Detectors such as GPTZero, Sapling AI Detector, GPT-2 Detector, DetectGPT, and GLTR become evident when applied to the detection of AI-generated code. Variants 2, 3, and 5 enable students to deceive most models, leading to a significant decrease in accuracy and TNR across the majority of models. These limitations stem from the fundamental differences between programming code and natural language text. Code adheres to strict rules, follows distinct patterns, and may not exhibit the linguistic characteristics that these models are designed to detect.
\end{framed}

\section{Discussion}
\subsection{Suggestions to SE and CS Educators}
AIGC has emerged as a transformative force in education, particularly in the domains of CS/SE. This section aims to distill insights from recent research papers, to provide educators with strategies and best practices for the successful integration of AI into CS/SE education.

Researchers from Hong Kong introduced the IDEE Framework, which serves as a guiding framework for utilizing generative AI in education \cite{su2023unlocking}. This framework emphasizes the importance of identifying desired outcomes, determining the appropriate level of automation, ensuring ethical considerations, and evaluating effectiveness. Moreover, the work by Kaplan et al. explores teachers' perspectives on generative AI technology and its potential implementation in education \cite{kaplan2023generative}. The authors suggest that teachers exhibit positive perspectives towards generative AI, with more frequent usage leading to increased positivity. Educators perceive generative AI as a tool for enhancing professional development and student learning. This suggest that embracing generative AI in education has the potential to positively impact both educators and students, fostering continuous growth and improved learning experiences.

On the other hand, the work by Chan et al. delves into the experiences and perceptions of Gen Z students and Gen X/Y teachers regarding the use of generative AI in higher education \cite{chan2023ai}. Students express optimism about the benefits of generative AI, while teachers emphasize the need for guidelines and policies to ensure responsible use. Furthermore, an AI Ecological Education Policy Framework for higher education \cite{chan2023comprehensive}. It addresses three dimensions: Pedagogical, Governance, and Operational, providing a comprehensive structure to navigate the implications of AI integration.

Based on the literature and research findings, it is imperative for educators to acknowledge the limitations inherent in current AIGC detectors, especially when dealing with code-based content. Our study emphasizes the critical need for ongoing exploration and advancement in specialized tools, algorithms, and frameworks. This becomes particularly evident when considering the predominant use of existing AIGC detectors for text-based content. Our results highlight a significant gap in their applicability to code-based AIGC, calling for the development of more refined tools and algorithms tailored to this specific context. This underscores the importance of staying abreast of technological advancements to ensure the effective detection and evaluation of AIGC within educational materials. In the absence of dedicated detectors tailored for code-based AIGC, we propose several key recommendations for educators:
{
\\
\vspace{0.2pt}
  \noindent  \textbf {1. Define Objectives:} Precisely outline the educational objectives that are in harmony with the application of generative AI. This step ensures a seamless integration of technology into educational purposes, fostering alignment between technology utilization and educational aspirations. For example, in a Python programming course, educators could utilize generative AI to enhance students' algorithmic creativity. Objectives might include collaborative design and optimization of sorting algorithms using AI assistance, fostering a deeper understanding of algorithmic principles.
\\
  \noindent  \textbf {2. Automation Level:} When incorporating generative AI into education, it is essential to carefully consider the level of automation to employ. This decision hinges on whether to pursue full automation, where AI systems handle educational tasks entirely on their own or opt for a supplementary approach that blends AI capabilities with human involvement. This choice plays a pivotal role in shaping the educational landscape, as it determines the extent to which technology should be seamlessly integrated into the learning process, aligning with the unique goals and requirements of each educational scenario. Take for instance a data science course, when integrating a code generation AI tool, consider the automation level. Decide whether to fully automate coding assignments, letting AI generate solutions, or adopt a supplementary approach, blending AI capabilities with human involvement. This choice influences how students learn programming, aligning with the curriculum's goals—whether to emphasize algorithmic logic through manual coding or leverage AI for rapid prototyping and problem-solving.
\\
  \noindent  \textbf {3. Ethical Focus: } Give paramount importance to ethical concerns. Develop comprehensive guidelines and policies to safeguard the responsible and ethical usage of AI in the educational context. Educators can, for example, develop guidelines to ensure responsible and ethical usage of AI algorithms in assignments, particularly those involving sensitive data. This may involve creating policies addressing issues such as data privacy, bias mitigation, and transparency, fostering a learning environment that upholds ethical standards in AI applications.
\\
  \noindent  \textbf {4. Continuous Evaluation: } It is essential to maintain an ongoing process of assessing the effectiveness of generative AI in education. This involves systematically monitoring how well the technology serves educational objectives, identifying areas for improvement, and adapting strategies as needed, ultimately contributing to improved outcomes and enhanced educational experiences. For example, in a Data Science course using generative AI for predictive modelling, maintain ongoing assessment. Systematically monitor how well the technology contributes to achieving predictive modelling objectives, identify areas for improvement, and adapt strategies as needed, ultimately enhancing students' data science skills.
\\
  \noindent  \textbf {5. Comprehensive Policies: } In the integration of generative AI within educational settings, it is imperative to develop comprehensive guidelines and policies based on empirical evidence and best practices. By adopting an evidence-based approach, educational institutions can ensure that their policies are not only robust and compliant but also rooted in real-world outcomes and experiences, safeguarding both the educational mission and the well-being of all stakeholders. In a Software Engineering curriculum incorporating a code generation AI tool, establish comprehensive policies. Develop guidelines based on empirical evidence and best practices to govern the use of AI-generated code in assignments. These policies may address code review processes, plagiarism detection, and the ethical implications of AI-assisted programming, ensuring a fair, transparent, and secure learning environment for both students and instructors.
\\
  \noindent  \textbf {6. Stay Informed: } Promote ongoing research and evaluation of AI integration in educational settings to stay informed about advancements, benefits, and risks associated with AI technology. For instance, in a university's Computer Science department, educators actively engage in continuous research on AI integration in coursework. This commitment ensures they stay informed about the latest advancements, benefits, and potential risks associated with AI technology, allowing for adaptive and optimized teaching methods based on the field's latest insights.
} 

\subsection{Threats to validity}
\subsubsection{Internal Validity}
The study faces challenges related to the varied prompts used to generate AIGC with ChatGPT. While these prompts aim to simulate average user inputs, they may not fully mirror real-world scenarios. Additionally, ChatGPT's non-deterministic nature, leading to diverse responses from the same prompt, could impact result reproducibility. 

Another challenge pertains to verifying the authenticity of source code written by humans. For instance, datasets sourced from platforms such as Kaggle could include content generated by LLM tools, blurring the distinction between human and AI contributions. However, since these datasets were released to the public before LLM models went mainstream, the impact is estimated to be minimal.

In specific scenarios, vague queries leading to responses such as "I'm sorry, as an AI language model, I am unable to provide code as the question lacked specific details about the expression to be evaluated" pose a risk to internal validity. Ambiguous or open-ended coding questions can result in inaccurate AIGC. Therefore, the datasets of the question prompts underwent preprocessing. During this process, rows containing responses such as the one mentioned were removed, ensuring the integrity and accuracy of the dataset used in our study.

\subsubsection{Construct Validity}
Construct validity explores the intricate interplay between theoretical constructs and empirical data, posing concerns about potential biases arising from platforms such as Kaggle, where datasets are curated and shared, and ChatGPT may have been trained on datasets similar to the one used in this study. However, understanding ChatGPT's ability to generate diverse and contextually relevant content, rather than simply replicate existing snippets, this research maintains objectivity and rigor. These insights allow for a nuanced approach, ensuring the study's authenticity and safeguarding the integrity of the research findings.

\subsubsection{External Validity}
It is crucial to acknowledge that the conclusions drawn in this study might be specific to the datasets analyzed. To enhance the generalizability of our study, we meticulously curated our dataset by gathering data from diverse sources, aiming to replicate real-world software development scenarios as closely as possible. Furthermore, we concentrated our efforts on the Python programming language, a deliberate choice made to maintain consistency and control within our study. 

\section{Related Work}
This section discusses several studies and research articles that explore the detection of AI-generated content, particularly in academic and educational contexts.

The study by Otterbacher \cite{otterbacher2023technical} highlights the need to develop a culture that promotes responsible and ethical use of generative AI in various domains, including science and education. Rather than relying solely on technical solutions to combat AI-generated content, the authors argue for a holistic approach that considers the broader implications of AI in academia. On the other hand, the work by Chaka \cite{chaka2023detecting} evaluated AI content detection tools, highlighting their limitations in accurately detecting AI-generated content, especially in academic contexts. These limitations can lead to academic integrity issues, including plagiarism. Chaka's study \cite{chaka2023detecting} also emphasizes the need for ongoing research to evaluate the accuracy and reliability of AI content detection tools. Improving these tools' ability to detect AI-generated content is crucial for combating AI-generated plagiarism in academia. AIGC, notably code produced by ChatGPT, shows promise in software-related tasks but raises concerns in education, particularly regarding plagiarism \cite{wang2023evaluating}.



The study by Adnan aimed to compare generated abstracts with original ones and assess the impact of instructions given to GPT-3.5 models on abstract quality \cite{altextual}. The result warns against misuse of results and emphasizes the risks in deploying ML tools in academia due to a 1\% false positive rate. At a larger scale, such misclassification could reject significant research, potentially impacting humanity.

Similar to many related studies suggesting AI’s inherent risks, Farrelly and Baker propose that AI, particularly LLM, holds considerable disruptive potential in education and society \cite{farrelly2023generative}. Minority and international students encounter increased allegations of academic integrity breaches due to these technologies. Nonetheless, they also offer valuable benefits, particularly for international students and individuals with disabilities.

In relation to our study, these works provide valuable insights into the limitations of existing AI content detection tools, especially in the context of AI-generated code. They underscore the need for a comprehensive evaluation of AIGC Detectors' performance and the development of ethical guidelines to ensure responsible AI-generated content use in education.

\section{Conclusion and Future Work}
The rise of generative AI models presents both opportunities and challenges, particularly in education and programming. Our study aimed to assess AI content detection tools' effectiveness in identifying AI-generated code, revealing their limitations and offering insights for educators and students. We examined various AI content detection models using a dataset containing human-written and AI-generated code. Our evaluation focused on metrics such as recall, precision, F1 score, accuracy, and AUC.

We have revealed significant sensitivity in GLTR, as indicated by its remarkably low p-value of 0.0296 in the Samples Paired t Test. This sensitivity led to notable accuracy discrepancies ranging from 0.4841 to 0.7693, exacerbated by specific code variants such as variants 2, 3, and 5. These findings underscore the imperative for immediate research efforts to enhance the reliability of AI-generated content detectors, safeguarding academic integrity in educational contexts. Addressing these challenges is pivotal for educators and institutions to adeptly navigate the complexities introduced by AI-generated code, ensuring the integrity of programming education.

Our findings suggest that these tools hold promise in distinguishing AIGC from human-written code but face challenges due to code complexity and writing style variations. Ethical guidelines for AIGC integration into education are essential. Moreover, we have found that GLTR is very sensitive to different variants introduced into the AIGC produced by ChatGPT, proven by the smallest p-value when the Samples Pair t-Test is applied to the AIGC Detector's accuracies and the wide range of the accuracy.

Educators and institutions should adopt strategies for responsible AI usage, considering curriculum design, assessment methods, and ethical directives. Additionally, the long-term impact of AIGC Detector tools on student skill development and creativity in programming warrants further exploration. In conclusion, our research highlights the evolving landscape of AI-generated code and the role of AIGC Detectors in education. Emphasizing responsible AI usage, ethical guidelines, and ongoing tool refinement can empower students in a technology-driven world while preserving academic integrity and fostering creativity. 


Future research should prioritize enhancing AI content detection models to effectively handle a wider range of code variations and writing styles, bolstering their reliability in educational contexts. Investigating the long-term impact of AIGC Detector tools on students' learning and creative engagement in programming can provide valuable insights for educators. Additionally, exploring the adaptability of AI-driven content detection models to architectural design, software design, and UML diagrams will contribute to a more comprehensive understanding of their applicability across diverse domains, fostering the development of improved educational tools and strategies.

\section{Data Availability}

The replication package, along with the associated data, has been made publicly available.\footnote{https://figshare.com/articles/dataset/Replication\_Package/24298036}
\balance
\bibliographystyle{ACM-Reference-Format}
\bibliography{Reference}

\end{document}